\documentstyle [12pt]{article}
 
\begin{document}
 
\baselineskip 7.5 mm

\begin{flushright}
\begin{tabular}{l}
CERN-TH/97-381 \\
hep-ph/9801212 
\end{tabular}
\end{flushright}
 
\vspace{8mm}

\begin{center}
 
{\Large \bf Star Wreck}

\vspace{8mm}
 
{\large
Alexander Kusenko,}$^{a,}$\footnote{ email address: Alexander.Kusenko@cern.ch}
{\large 
Mikhail Shaposhnikov,}$^{a,}$\footnote{ email address:
  mshaposh@nxth04.cern.ch} 
{\large P.~G.~Tinyakov,}$^{b,}$\footnote{email address: 
peter@flint.inr.ac.ru} \\
{\large and 
Igor I. Tkachev}$^{b,c,}$\footnote{ email address:
tkachev@physics.purdue.edu } \\
 
\vspace{6mm}
$^a$Theory Division, CERN, CH-1211 Geneva 23, Switzerland \\
$^b$Institute for Nuclear Research, 60th October Anniversary Prospect 7a,
Moscow, Russia 117312 \\
$^c$Department of Physics, Purdue University, West Lafayette, IN 47907, 
USA \\

\vspace{12mm} 
 
{\bf Abstract}

\end{center}
Electroweak models with low-energy supersymmetry breaking
predict the existence
of stable non-topological solitons, Q-balls, that can be 
produced in the early universe.  
The relic Q-balls can accumulate inside a neutron star and gradually absorb
the baryons into the scalar condensate.  This causes a slow reduction in the
mass of the star.   When the mass reaches a critical value, the neutron
star becomes unstable and explodes.  The cataclysmic destruction of the
distant neutron stars may be the origin of  the gamma-ray bursts.  
 
\vspace{8mm}
 
\begin{flushleft}
\begin{tabular}{l}
CERN-TH/97-381 \\
December, 1997
\end{tabular}
\end{flushleft}

\vfill
 
\pagestyle{empty}
 
\pagebreak
 
\pagestyle{plain}
\pagenumbering{arabic}

The gravitational contraction transforms 
the low-density gaseous baryonic matter into the dense bodies, stars, 
which, in turn, burn out leaving behind the remnants of even higher 
density: white dwarfs, neutron stars, and black holes.  Isolated cold white
dwarfs and neutron stars do not evolve any further if the particle content
of Nature comprises that of the Standard Model. As we shall see, however,
supersymmetry can change the destiny of baryons in the universe.

In supersymmetric theories, the presence of squarks
makes it possible for the baryon number to be carried by
scalar particles.  Since the squarks are supposed to be much heavier than
the quarks to satisfy the present experimental limits,  
one might think that they cannot play a role in the late stellar evolution.
However, it was recently shown \cite{ks,dks} that in a class of SUSY
theories with low-energy supersymmetry breaking the lowest-energy state
with a given (and sufficiently large) baryon number is 
a coherent state of squarks, Q-ball, rather than the ordinary nuclear
matter.  A large stable Q-ball, of the kind that may have been copiously
produced in the early universe~\cite{ks} and can presently exist as a form
of dark matter~\cite{ks}, can grow by absorbing the matter nuclei and
annexing their baryon number to its scalar condensate~\cite{dks,kkst}. It
is of interest, therefore, whether the presence of the relic Q-balls can
alter the ultimate fate of the baryonic astrophysical objects.

In this Letter we show that the relic Q-balls can accumulate in
both the neutron stars and the white dwarfs, and cause their eventual 
destruction.  While a Q-ball consumes the baryon number of a neutron star, 
the energy liberated in the process is emitted, predominantly, in
neutrinos.  The dramatic weight loss associated with this process brings
the masses of both the neutron stars and the white dwarfs to their
respective minimal values beyond which they are unstable.  This leads to
an explosion.  The lifetime of a contaminated star  is determined by the
parameters  
of the underlying supersymmetric theory and can be in the Gyr range. The
energy released from the cataclysmic death of a neutron star may be the
origin of the observed gamma-ray bursts (GRB).   Assuming a cosmological
distance scale, and using the observational data on GRB, one can put a
constraint on the  Q-ball parameters and the mode of supersymmetry breaking.   
Another constraint, a lower limit on the neutron star lifetime, comes from
the observation of pulsars, rotating magnetized neutron stars, with ages
over 0.1 Gyr.  

In general, Q-balls are non-topological solitons~\cite{fls,coleman,lp} 
that owe their stability to the conservation of some global charge.  
Regardless of the way supersymmetry is broken, the Minimal Supersymmetric
Standard Model (MSSM) admits Q-balls~\cite{ak_mssm},
held together by the tri-linear couplings in the scalar potential that 
originate from the Yukawa couplings in the superpotential, and also from the
soft supersymmetry-breaking terms.   Such Q-balls, whose charges are 
arbitrary~\cite{ak_qb}, could be produced in the early universe~\cite{ak_pt}
but would have eventually decayed into the light fermions that carry the same
global charges.  Formation and subsequent decay of the unstable Q-balls
could have various cosmological consequences, in particular, with respect
to the baryogenesis~\cite{ks,em}. 

If, however, the scalar potential $U(\phi)$ has a flat direction\footnote{
Flat potentials arise naturally in theories with
low-energy supersymmetry breaking (see, {\it e.\,g.}, Refs.~\cite{gia1,gmm})
and are generic for gauge-mediated scenarios (for a
review see, {\it e.\,g.}, Ref. \cite{gauge-mediated}).}
and $U(\phi) \sim m^4=const$ for large $\phi$, then the mass of 
a soliton with charge $Q_{_B}$ is $M_{_Q}  
\propto Q_{_B}^{3/4}$ for
large $Q_{_B}$~\cite{dks,ks}.  Therefore, for arbitrary $m$, the mass of
a baryonic Q-ball with a sufficiently large charge $Q_{_B}$ is less than
$m_p Q_{_B}$.  Such a Q-ball is stable because it cannot decay into
the ordinary matter baryons.  The natural values of the mass parameter $m$ in
theories with low-energy supersymmetry lie between 100~GeV and 100~TeV.
Stable baryonic Q-balls can be produced in the early universe from the
breakdown of a coherent scalar condensate~\cite{ks} and can contribute to
dark matter.  The primordial scalar condensate is invoked in the
Affleck--Dine scenario for baryogenesis~\cite{ad}.  It is possible,
therefore, that the baryonic matter in the universe and  the solitonic dark
matter may share the same origin.  

We assume the condensate inside a Q-ball to be a combination of 
squarks (and, possibly, some other, non-baryonic, scalar fields, like
sleptons and the Higgs fields) that transforms as a singlet under the
action of the gauge group.  The latter is a necessary condition for 
the scalar field to be a flat direction of the potential in a  
supersymmetric theory.  At the same time, this allows one to describe a
Q-ball semi-classically as a field configuration in which all the gauge
charges vanish~\cite{kst}.  Following 
Refs.~\cite{dks,ks,kkst}, we will assume that the mass of a soliton with
charge (baryon number) $Q_{_B}$ is  $M_{_Q}  \simeq (4\pi \sqrt{2}/3) \, m
\, Q_{_B}^{3/4}$, its radius is  $R_{_Q}  \simeq  (1/\sqrt{2})\,  m^{-1} \,
Q_{_B}^{1/4}$, and the maximal scalar VEV inside is $\phi_{_Q}  \simeq
(1/\sqrt{2}) \,  m \, Q_{_B}^{1/4}$.

The dark-matter Q-balls, with expected baryon charges over
$10^{21}$~\cite{kkst}, are too heavy to stop in the Earth or the Sun. 
However, they do stop and accumulate inside a neutron star.  The equation
of motion for a small heavy object moving with friction inside a
gravitating body is

\begin{equation}
\ddot{R}=-\Omega^2 R - \gamma \dot{R},
\label{eq_fall}
\end{equation}
where $R$ is the distance to the center of the star. For simplicity,
we work in the Newtonian approximation to gravity and  neglect the effects 
of general relativity. Assuming that the Q-ball has a geometrical 
cross-section of scattering, we obtain 

\begin{eqnarray}
\Omega & = & \sqrt{\frac{4 \pi}{3} \frac{\rho}{M_{_P}^2}}, 
\label{Omega}\\
\gamma & \sim & \frac{\rho v_n}{m^3 Q_{_B}^{1/4}}, 
\label{gamma}
\end{eqnarray}
where $\rho $ is the matter density (assumed to be constant inside the
star), $v_n$ is the average nucleon velocity in matter, and $M_{_P}$ is the
Planck mass.  Depending on the matter density, the Q-ball
can oscillate (if $\Omega \gg \gamma$), or it can wade to the center
gradually (if $\Omega \ll \gamma$).  For the neutron star density, the
latter is the case because $\gamma/\Omega \sim 10^2 \, (10^{24}/Q_{_B})^{1/4}
\gg 1$. The Q-balls settle in the center of the star in a matter of seconds. 

The interactions of a large baryonic soliton with matter proceeds in two
stages.  First, an incoming nucleon (or a nucleus) collides with a Q-ball 
and dissociates into quarks.  This is because QCD deconfinement takes place
inside every baryonic Q-ball due to the spontaneous SU(3) breaking 
by the squark VEV's.  This process is accompanied by the emission of,
roughly, 1 GeV in pions and is the basis for the experimental detection of
the relic Q-balls discussed in Ref. \cite{kkst}.  Second, the quarks inside
a Q-ball transfer their baryonic charge to the Q-ball.  As a result, the
Q-ball grows in baryon number and in size at the expense of the incoming
baryons.  The rate of ``processing'' of quarks inside a Q-ball is determined
by the physics beyond the Standard Model, which we will assume to be
described by the MSSM.   

In the ordinary matter, the rate of nucleon collisions with a large Q-ball 
is much lower than the time scale associated with the conversion of
quarks into the quanta of the scalar condensate~\cite{kkst}.  In contrast,
a Q-ball inside a neutron star receives nucleons at a much higher rate,
which may exceed the rate of ``processing''.  In this case, the growth of
the baryonic condensate is determined by the cross-sections of the MSSM
reactions that generally proceed through a gluino exchange.  
The VEV of squarks inside a Q-ball causes the mixing of quarks and gluinos,  
through which they gain masses.  Therefore, the rate of a quark decay
inside a Q-ball is suppressed by the magnitude of the VEV and takes place
predominantly in the outer region of the soliton, where the scalar VEV is
small.  Hence, the rate of change in the soliton charge $Q_{B}$ is
proportional to the surface area $4\pi R_{_Q}\propto Q_{_B}^{1/2}$ of a Q-ball:
$dQ_{B}/dt = \alpha' Q_{_B}^{1/2}$.  If the quarks can be absorbed
into the condensate faster than the rate at which they enter the Q-ball,
the growth of the soliton's baryon number is also proportional to the
surface area: $dQ_{B}/dt = \alpha'' Q_{_B}^{1/2}$, where $\alpha''$
is determined by the flux of incoming baryons.  Therefore, 
for an individual Q-ball in matter, 

\begin{equation}
dQ_{B}/dt = \alpha Q_{_B}^{1/2}, 
\label{one_q}
\end{equation}
where $\alpha=min(\alpha',\alpha'')$. 

Unlike the Grand Unified monopoles, which catalyze the proton
decay~\cite{cr} but do not grow in size, Q-balls convert the baryon number
of the nucleons into that of the condensate.  As a result, their ability to
absorb nuclei grows with their surface area, in accordance with equation
(\ref{one_q}). This makes the behavior of the Q-balls inside the star very
different from that of the monopoles~\cite{kr}.  

Some Q-balls may be impervious for the electrons~\cite{kkst}.
This causes an additional suppression of their interactions with the ordinary
matter because,  after the first proton is absorbed,  the Coulomb barrier
can prevent the other nuclei from approaching a Q-ball.  Depending on their
ability to retain the electric charge, Q-balls can be separated in two
classes: Supersymmetric Electrically Neutral Solitons  (SENS) and
Supersymmetric Electrically Charged Solitons (SECS)~\cite{kkst}.  In a
neutron star, where all nucleons are electrically neutral, there is no
difference between SENS and SECS.  However, when we discuss the fate of
the white dwarfs, we will have to make a distinction between the two types. 

When a neutron star is born in a supernova explosion, the first relic
Q-balls can be slowed down by the dense nuclear matter.  They drift towards
the center.  As they grow in size at the expense of the ambient matter,
they coalesce and consolidate into a single Q-ball with a very large baryon
number.  From that point on, the baryon number of a neutron star is
consumed by the Q-balls at the rate 

\begin{equation}
\frac{dQ_{B}}{dt} = \frac{q_*}{\tau} + \alpha Q_{_B}^{1/2}, 
\label{dqdt}
\end{equation}
where $q_*$ is the average charge of a newly arriving Q-ball at the time it
merges with the large central Q-ball, and $\tau$ is the time scale
associated with the flux of Q-balls.  If some Q-balls are already present
in the neutron star at the time of its formation, this is accounted for by
the appropriate initial conditions.  (This may be relevant, in particular,
to the Population III stars that may form prior to the formation of the 
galaxies and may be accompanied by a cloud of primordial Q-balls from the
start.)  

In the case of a neutron star, the very high matter density ensures a
plentiful supply of nucleons.  Parameter $\alpha$ depends, therefore, 
entirely on the rate of absorption of quarks into the condensate.  This
process can be thought of as a quark-quark annihilation in the external
background scalar field $e^{i\omega t} \phi(x)$.  The cross-section
$\sigma $ of this process depends on the parameters of the MSSM and the
structure of a given Q-ball.  We parameterize $\sigma$ as 
$\sigma =  \beta/m^2$. 
As was mentioned above, the absorption of quarks into the condensate can
take place in the outer region of the soliton, where the gluino mass is not
too large.  The effective volume of this region can be estimated using the
form of $\phi(x)$ from Ref.~\cite{dks} and is proportional to $\omega^{-3} 
\propto m^{-3}$.  Therefore, $\alpha \propto \sigma \omega^{-3} \propto 
m^{-5}$.  In what follows we will use $\alpha$ as a phenomenological
parameter, 

\begin{equation}
\alpha = 10^8 \, \left ( \frac{m}{1 {\rm TeV}} \right )^{-5} \, 
\beta \, {\rm s}^{-1}.
\label{alpha}
\end{equation}

The fate of a neutron star is sealed when it captures the first Q-ball.
In less than a second it reaches the central region of the star and begins
to grow. In the process it may coalesce with the other solitons captured at
later times. When the baryon number of the central
Q-ball exceeds $\sim (q_*/\alpha \tau)^2$, 
the solution of equation (\ref{dqdt}) is 
$Q_{_B}(t) \approx (\alpha t/2)^2$.  The lifetime of a 
neutrons star depends on $m$ and on the cross-section of $qq$ annihilation,  

\begin{equation}
t_s \sim \frac{1}{\beta} \times 
\left ( \frac{m}{200\, {\rm GeV}} \right )^{5} {\rm Gyr}, 
\label{lifetime}
\end{equation}
and can be as low as $10^{-2}$ Gyr for $m \sim 100$ GeV, or can exceed 
10 Gyr for $m \stackrel{>}{_{\scriptstyle \sim}} 300$ GeV,
if $\beta \sim 1$. Since the
pulsars with ages around 0.1 Gyr are known to exist, there is a lower bound
on $t_s$ and, correspondingly, on $m^5/\beta$. 
If $\beta \ll 1$ and $m$ is in the TeV region then
the lifetime of a neutron star exceeds the age of the
Universe and the relic Q-balls play no role in the stellar evolution at 
present time. 

The lifetime of the star has little  dependence on the flux of Q-balls in a
wide range of parameters.  If the baryonic solitons make up a substantial
fraction of dark matter ($\Omega_{_Q} \sim 1$), their flux is estimated to
be  $F= 1.2 \times 10^2 Q^{-3/4} (m/1 \, {\rm TeV})^{-1} $ cm$^{-2}$
s$^{-1}$ sr$^{-1}$~\cite{kkst}.  If, however, $\Omega_{_Q} \ll 1$, then 
it is possible that the probability for a neutron star to capture a Q-ball
is very small.  In this case, the average lifetime of a neutron star can 
be much greater than $t_s$ in equation (\ref{lifetime}). 

The energy released from the consumption of nuclear matter by the 
Q-balls is 

\begin{equation}
\frac{dE}{dt} = m_n \frac{dQ_{_B}}{dt} \sim m_n \alpha^2 t. 
\label{dedt}
\end{equation}
For $\alpha \sim 10^8$ s$^{-1}$ and $t \sim 1$ Gyr, 
this yields $dE/dt \sim 10^{30}$ erg/s.  This energy is radiated
away in thermal neutrinos and photons, with no observational consequences. 
Meanwhile, the star slowly looses its mass.

Neutron stars are stable in a certain range of masses.  Assuming
Harrison--Wheeler equation of state, one obtains a lower limit on the mass  
$M_{min} \approx 0.18 M_{\odot}$~\cite{st}.  By the time the Q-balls annex 
80\% of the neutron star baryon number, a similar fraction of the star's
mass is radiated away with thermal neutrinos.  At this point the mass of
the Q-ball inside the star, $M_{_Q} \sim 10^{-11} M_{\odot}$, is still 
too small to affect the conditions of hydrostatic equilibrium.   
When the mass of a neutron 
star becomes less than $M_{min}$, its gravity is no longer sufficient to
support the stability of nuclear matter.  The result is an explosion, in
which the nuclear matter undergoes a transformation into protons and
electrons.  The explosion is associated
with the energy release $\sim 10^{52}...10^{53}$ erg, almost as grandiose
as that of a supernova.  As the gravity subsides at the onset of the
explosion, the unleashed neutrons can decay into protons and electrons with
an emission of neutrinos and gamma-rays.   

The emission of gamma-rays may account for the observed gamma-ray
bursts.  The energy of the explosion corresponds to that inferred from the
data on gamma-ray bursts under the assumption that they take place at
the redshift $z=1-2$~\cite{mcl}.
The lifetime of a neutron star we obtained in equation (\ref{lifetime}) 
depends sharply on the parameter $m$ determined by the physics beyond the
Standard Model.  It seems natural for $t_s$ to be of order a few billion
years.  If this is the case, there is a population of dying neutron stars
at the redshift of a few that can be the source of gamma-ray bursts.  An
exploding neutron star with subcritical mass would produce a fireball of
protons, electrons and decaying neutrons.  The study of such explosion, to
which some of the discussion in Ref.~\cite{mcl} may apply, lies outside the
scope of the present work.   

A similar analyses of white dwarfs shows that they can also
capture the relic Q-balls.  However, since $\Omega \gg \gamma$, equation
(\ref{eq_fall}) yields an oscillatory solution.  After a number of
oscillations, the Q-balls settle down in the center of a white dwarf.  If
they are of the SECS type, their effect is negligible.  If, however, they
are SENS,  they consume the white dwarf from the inside, just like in the
case of a neutron star.  The white dwarfs also become unstable when their
masses become less than $2.2 \times 10^{-3}\, M_\odot$, at which point they
also explode. However, the energy liberated by the explosion is, probably,
too small to cause any significant observational consequences.  

In summary, the relic Q-balls can accumulate inside a neutron star and
cause a gradual reduction in the mass of the star until it reaches a
critical value.  At that point gravity can no longer support the stability
of the nuclear matter, and an explosion follows.  The lifetime of a neutron
star depends on physics beyond the Standard Model, but can be in the Gyr
range.  If so, the explosions of distant neutron stars can be the origin of
the gamma-ray bursts. 

We thank V.~A.~Kuzmin for many stimulating discussions.  P.~G.~T. and
I.~I.~T. thank Theory Division at CERN for hospitality.

\end{document}